\documentclass[a4paper,fleqn,usenatbib]{aa}

\usepackage[T1]{fontenc}
\usepackage{ae,aecompl}

\usepackage{graphicx}   
\usepackage{amsmath}    
\usepackage{amssymb}    
\usepackage{mathtools}
\usepackage{comment}
\usepackage{subfigure}
\usepackage{hyperref}
\hypersetup{
    colorlinks = true,
    citecolor = blue,
}

\graphicspath{ {./images/} }

\defcitealias{soldateschi_2020}{SBD20}

\begin{document}

\title{Magnetic deformation of neutron stars in scalar-tensor theories}

\author{J. Soldateschi
          \inst{1,2,3}
          \and
          N. Bucciantini\inst{2,1,3}
          \and
          {L.} {Del Zanna}\inst{1,2,3}
          }

        \offprints{J. Soldateschi, \email{jacopo.soldateschi@inaf.it} or N. Bucciantini, \email{niccolo.bucciantini@inaf.it}}

   \institute{Dipartimento di Fisica e Astronomia, Universit\`a degli Studi di Firenze, Via G. Sansone 1, I-50019 Sesto F. no (Firenze), Italy
         \and
             INAF - Osservatorio Astrofisico di Arcetri, Largo E. Fermi 5, I-50125 Firenze, Italy
                \and
                        INFN - Sezione di Firenze, Via G. Sansone 1, I-50019 Sesto F. no (Firenze), Italy
             }
    
\date{Received XXX; accepted YYY}

\abstract{
Scalar-tensor theories are among the most promising alternatives to general relativity that have been developed to account for some long-standing issues in our understanding of gravity. Some of these theories predict the existence of a non-linear phenomenon that is spontaneous scalarisation, which can lead to the appearance of sizable modifications to general relativity in the presence of compact matter distributions, namely neutron stars. On the one hand, one of the effects of the scalar field is to modify the emission of gravitational waves that are due to both variations in the quadrupolar deformation of the star and the presence of additional modes of emission. On the other hand, neutron stars are known to harbour extremely powerful magnetic fields which can affect their structure and shape, leading, in turn, to the emission of gravitational waves -- in this case due to a magnetic quadrupolar deformation. In this work, we investigate how the presence of spontaneous scalarisation can affect the magnetic deformation of neutron stars and their emission of quadrupolar gravitational waves, both of tensor and scalar nature. We show that it is possible to provide simple parametrisations of the magnetic deformation and gravitational wave power of neutron stars in terms of their baryonic mass, circumferential radius, and scalar charge, while also demonstrating that a universal scaling exists  independently of the magnetic field geometry and of the parameters of the scalar-tensor theory. Finally, we comment on the observability of the deviations in the strain on gravitational waves from general relativity by current and future observatories.
}

\maketitle



\section{Introduction}
\label{sec:intro}

Scalar-tensor theories (STTs) are among the most studied alternatives to general relativity (GR). Since the pioneering paper by \citet{brans_machs_1961}, much work has been devoted to study STTs and their phenomenology \citep{matsuda_hydrodynamic_1973,novak_spherical_1998,fujii_scalar-tensor_2003,zhang_2019}. The main premise is taking a scalar field, which is non-minimally coupled to the metric and which plays the role of an effective space-time dependent gravitational `constant', and adding it to the gravitational action. The effects of the scalar field manifest themselves in the presence of matter, which acts as a source for the scalar field. They range from cosmological scales, where the non-minimal coupling can account for cosmological observations without needing to resort to the existence of a dark sector (see e.g. \citealt{capozziello_extended_2011}), up to the small scales of compact objects. Unfortunately, black holes are not useful in constraining STTs because the no-hair theorem also holds  in such theories, preventing them from developing a scalar charge \citep{hawking_black_1972,berti_testing_2015}. This means that neutron stars (NSs) hold a special importance in testing GR -- and even more so given that their compactness allows the scalar field to manifest its effects in a non-perturbative fashion through spontaneous scalarisation \citep{damour_nonperturbative_1993}, while allowing the tight observational constraints in the weak-gravity regime to be fullfilled \citep{shao_constraining_2017}. These effects are of a varied nature, including the emission of additional modes of gravitational waves (GWs) \citep{pang_2020}, a modified relation between the NS mass and radius and its central density, a modification of NS merger dynamics \citep{shibata_coalescence_2014}, as well as a variation in the frequency of normal modes of NSs \citep{sotani_2005} in their tidal and rotational deformation \citep{pani_2014,doneva_iq_2014} and their light propagation \citep{bucciantini_2020}.
\\\\
There are various effects that can induce quadrupolar modifications in the shape of NSs. For example, crustal deformations could lead to the formation of mountains on the surface of the NS \citep{haskell_2006}, which would then cause the emission of continuous GWs (CGWs) \citep{usho_2000}. Mountains are also expected to form during accretion in a process called magnetic burial \citep{melatos_2005}. GWs are expected to be also released during starquakes, following a rearrangement of the magnetic field of the star, which may excite some of its oscillation modes \citep{keer_2015}. Moreover, oscillation modes such as the axial r-mode are unstable due to the emission of GWs \citep{andersson_1998}. Finally, if the magnetic axis of a NS is not aligned to its rotation axis, the star acquires a time-varying quadrupolar deformation which leads to the emission of CGWs, and it is believed that a NS endowed with a strong toroidal magnetic field will develop an instability which flips the star to an orthogonal rotator, maximising its emission of CGWs \citep{cutler_2002}.  This is particularly relevant in the context of newly born and rapidly rotating magnetars, which have been invoked as possible engines for long and short gamma-ray bursts \citep{Dall'Osso_Shore+09a,Metzger_Giannios+11a,rowlinson_2013}, where GWs can compete with electromagnetic losses and substantially modify the energetic budget and evolution of these systems. It is clear that the intense magnetic fields stored in NSs have a great importance in determining their GW phenomenology.
\\\\
Neutron stars are known to harbour extremely powerful magnetic fields, in the range $10^{8-12}$G for normal pulsars and up to $10^{16}$G for magnetars (while newly born proto-NSs are thought to contain magnetic fields as high as $10^{17-18}$G, see e.g. \citealt{del_zanna_chiral_2018,ciolfi_2019,franceschetti_2020}). It is important in this sense to study the interplay between the magnetic and the scalar field in shaping the quadrupolar deformation of the NS, even more so because of its connection to the emission of GWs. Moreover, this is relevant to the study of how the presence of an additional channel for the emission of quadrupolar waves - that of `scalar waves' - affects the overall emission of quadrupolar GWs, establishing the extent to which the emission of scalar waves competes with the tensor one. In this paper, we build upon  \citet{soldateschi_2020} (hereafter \citetalias{soldateschi_2020}), where we studied the general problem of axisymmetric models of NSs in STTs in the presence of spontaneous scalarisation to investigate the magnetic deformation of NSs in a class of STTs containing spontaneous scalarisation in light of GW emissions. In \citetalias{soldateschi_2020}, we showed that the scalar field is expected to modify the magnetic deformation of NSs, but we investigated just a few select configurations for a single STT. Here we investigate the full parameter space. We refer to \citetalias{soldateschi_2020} for the details of the code and formalism we used and for the definition of the main physical quantities.
\\\\
 In Sect.~\ref{sec:equations}, we briefly introduce the problem of magnetohydrodynamics (MHD) in STTs. In Sect.~\ref{sec:results}, we show our results regarding the magnetic deformation of NSs in STTs, also describing  the consequences for the emission of gravitational and scalar radiation. Finally, we present our conclusions in Sect.~\ref{sec:conclusions}.

\section{Scalar-tensor theories in a nutshell}\label{sec:equations}

In the following, we assume a signature $\{-, +, +, +\}$ for the
spacetime metric and use Greek letters $\mu$, $\nu$, $\lambda$, ... (running from
0 to 3) for 4D spacetime tensor components, while Latin letters
$i$, $j$, $k$, ... (running from 1 to 3) are employed for 3D spatial tensor components. Moreover, we use the dimensionless units where $c = G = \mathrm{M}_\odot = 1$, and we absorb the $\sqrt{4\pi}$ factors in the definition of the electromagnetic quantities. All quantities calculated in the Einstein frame (E-frame) are denoted with a bar ($\bar{\cdot}$) while all quantities calculated in the Jordan frame (J-frame) are denoted with a tilde ($\tilde{\cdot}$).
\\\\
The action of STTs in the J-frame, according to the `Bergmann-Wagoner formulation' \citep{bergmann_comments_1968,wagoner_scalar-tensor_1970,santiago_2000}, is
\begin{equation}
\begin{split} \label{eq:joract}
        \tilde{S}&= \frac{1}{16\pi}\int \mathrm{d}^4x \sqrt{-\tilde{g}}\left[ \varphi \tilde{R} - \frac{\omega (\varphi)}{\varphi} \tilde{\nabla} _\mu \varphi \tilde{\nabla} ^\mu \varphi - U(\varphi) \right] +\\
        &+ \tilde{S}_\mathrm{p}\left[ \tilde{\Psi} , \tilde{g}_{\mu \nu} \right]  \; ,
\end{split}
\end{equation}
where $\tilde{g}$ is the determinant of the spacetime metric $\tilde{g}_{\mu \nu}$, $\tilde{\nabla} _\mu$ its associated covariant derivative, $\tilde{R}$ its Ricci scalar, while $\omega (\varphi)$ and $U(\varphi)$ are, respectively, the coupling function and the potential of the scalar field $\varphi$, and $\tilde{S}_\mathrm{p}$ is the action of the physical fields $\tilde{\Psi}$. In the E-frame, the action is obtained by making the conformal transformation $\bar{g}_{\mu \nu}= \mathcal{A}^{-2}(\chi) \tilde{g}_{\mu \nu}$, where $\mathcal{A}^{-2}(\chi) =\varphi (\chi)$ and $\chi$ is a redefinition of the scalar field in the E-frame, related to $\varphi$ by
\begin{equation}
\frac{{\mathrm d}\chi}{{\mathrm d}\ln \varphi} = \sqrt{\frac{\omega{(\varphi)} +3}{4}} \; .
\end{equation}
In the following, we narrow the focus down to the simpler case of a massless scalar field, thus, $U(\varphi)=0$. In the E-frame, the scalar field is minimally coupled to the metric. This means that Einstein's field equations retain their usual form in the E-frame, taking  into account the fact that the energy-momentum tensor is now the sum of the physical one and of the scalar field one. Instead, Eq.~\ref{eq:joract} shows that the scalar field is minimally coupled to the physical fields in the J-frame. This implies that MHD equations in the J-frame have the same expression as in GR. In addition to the metric and MHD equations, in STTs we have an additional equation to solve for the scalar field. In the Einstein frame, it reads:
\begin{equation}\label{eq:scal}
        \bar{\nabla} _\mu \bar{\nabla} ^\mu\chi = -4\pi \alpha _\mathrm{s} \bar{T}_{\mathrm{p}} \; ,
\end{equation}
where $\bar{\nabla} _\mu$ is the covariant derivative associated to the E-frame metric $\bar{g}_{\mu \nu}$, $\bar{T}_{\mathrm{p}}= \bar{g}_{\mu \nu}\bar{T}_{\mathrm{p}}^{\mu \nu}$, $\bar{T}^{\mu \nu}_{\mathrm{p}}$ is the physical energy-momentum tensor in the E-frame and $\alpha _\mathrm{s}(\chi)~=~d \ln \mathcal{A}(\chi) / d\chi $.

\section{Results}\label{sec:results}
In the following, we focus on the case of static NSs in the weak magnetic field regime, meaning that the effects induced by the magnetic field on the deformation of the star are well-approximated by a perturbative approach; this was shown to be the case for $\tilde{B}_\mathrm{max}\lesssim 10^{17}$G \citep{pili_general_2015,bucciantini_role_2015}. This is much less than the critical field strength, of the order of $10^{19}$G, set by the energy associated to the characteristic NS density \citep{lattimer_neutron_2007}. Moreover, we focus only on the mass range of stable configurations.

The Newtonian quadrupole deformation $\bar{e}$ of a NS in STTs is formally defined as 
\begin{equation}\label{eq:deform}
    \bar{e} = \frac{\bar{I}_{zz}-\bar{I}_{xx}}{\bar{I}_{zz}} \quad ,
\end{equation}
where $\bar{I}_{zz}$ and $\bar{I}_{xx}$ are the Newtonian moments of inertia in the E-frame, accounting for both the physical and scalar fields energy density (see App.~C of \citetalias{soldateschi_2020}). This definition has the advantage that it is given as an integral over the star.
As is already known in Newtonian gravity \citep{wentzel_1960,ostriker_1969} and in GR \citep{frieben_equilibrium_2012,pili_general_2017}, in the limit of weak magnetic fields and slow rotation rates, the quadrupole deformation can be expressed as a bilinear combination of $B^2_\mathrm{max}$, where $B_\mathrm{max}=\max [\sqrt{B_i B^i}],$ with $B$ as the NS magnetic field, and its rotation rate  \citep{pili_general_2017}. Equivalently, instead of using $B^2_\mathrm{max}$ one can parametrise the quadrupole deformation also in terms of $\mathcal{H}/W$, where $\mathcal{H}$ is the magnetic energy of the NS, defined in the J-frame as
\begin{equation}
\tilde{\mathcal{H}} = \pi \int \mathcal{A}^3\tilde{B}_i \tilde{B}^i \sqrt{\bar{\gamma}}\mathrm{d}r \mathrm{d}\theta \; ,
\end{equation}
and with $W$ as its binding energy (which in STTs is properly defined in the E-frame). The true gravitational quadrupole moment is properly defined from the asymptotic structure of the metric terms \citep{bonazzola_1996,gourg_rel_stars_2010,doneva_iq_2014}, while the moment of inertia is only properly defined for rotators; however, it has been found that the Netwonian approximation is quite reliable \citep{pili_general_2015}. We note here that Eq.~A2 in \citep{pili_general_2015} is not formally correct, because it neglects frame dragging, while it can be shown that, for compact systems like NSs, this contributes about 10-15\% to the moment of inertia.

In our STT scenario, we found that $\bar{e}$ still follows a linear trend with $\tilde{B}^2_\mathrm{max}$ (or $\tilde{\mathcal{H}}/\bar{W}$), although with coefficients that bear a potentially much stronger dependence on  the baryonic mass $M_0$ (defined as in Eq.~C.4 in \citetalias{soldateschi_2020}) than in GR, depending on the value of the parameter regulating spontaneous scalarisation, $\beta _0$ (see below for its definition). In particular, in the limit $\tilde{B}_\mathrm{max} \rightarrow 0$, keeping fixed $M_0$ and $\beta _0$:
\begin{equation}\label{eq:coeffs}
        |\bar{e}| = c_\mathrm{B} \tilde{B}^2_\mathrm{max} + \mathcal{O}\left( \tilde{B}^4_\mathrm{max} \right) , \quad |\bar{e}| = c_\mathrm{H} \frac{\tilde{\mathcal H}}{\bar{ W}} + \mathcal{O}\left( \frac{\tilde{\mathcal H} ^2}{\bar{ W} ^2} \right),
\end{equation}
where $c_\mathrm{B}=c_\mathrm{B}(M_0,\beta _0)$ and $c_\mathrm{H}=c_\mathrm{H}(M_0,\beta _0)$ are the `distortion coefficients', and  $\tilde{B}_\mathrm{max}$ is normalised to $10^{18}$G. 

In order to compute the distortion coefficients of NSs in the low magnetic field regime, we computed several numerical models of magnetised NSs with stronger magnetic fields, and then we interpolated the results according to the functional form of Eq.~\ref{eq:coeffs}. We studied only configurations belonging to the stable branch of the mass-density diagram, that is with masses and central densities lower than that of the configuration with maximum mass. Following \citetalias{soldateschi_2020}, in the 3+1 formalism for spacetime splitting we  used the \texttt{XNS} code \citep{bucciantini_general_2011,pili_axisymmetric_2014} to solve numerically the equations for static and axisymmetric equilibrium configurations with a magnetic field, in the XCFC (eXtended Conformally Flat Condition) approximation \citep{cordero-carrion_improved_2009}. By assuming that the spatial metric is conformally flat, this approximation allows us to cast the Einstein equations in a simplified, decoupled and numerically stable form that can be solved hierarchically. Even if it is not formally exact, the XCFC approximation has proved to be highly accurate for rotating NSs \citep{iosif_2014,Camelio_Dietrich+19a}. We used a 2D grid in spherical coordinates extending over the range $r=[0,100]$ in dimensionless units, corresponding to a range of $\sim$150 km, and $\theta=[0,\pi]$.  The grid has 400 points in the $r$-direction, with the first 200 points equally spaced and covering the range $r=[0,20]$, and the remaining 200 points logarithmically spaced ($\Delta r_i/\Delta r_{i-1} = \mathrm{const}$), and 200 equally spaced point in the angular direction. We adopted a polytropic equation of state (EoS) $\tilde{p}=K_{\mathrm a} \tilde{\rho}^{\gamma_{\mathrm a}}$, where $\tilde{p}$ and $\tilde{\rho}$ are the pressure and rest mass density of the fluid, respectively, with $\gamma_{\mathrm a}=2$  and $K_{\mathrm a}=110$ in dimensionless units. This has already been used by several authors \citep{bocquet_rotating_1995,kiuchi_relativistic_2008,frieben_equilibrium_2012,pili_axisymmetric_2014,soldateschi_2020} as an approximation of more complex and physically motivated EoSs \citep{lattimer_neutron_2007,baym_2018}, above nuclear densities. We chose to solve the metric and scalar field equations in the E-frame and the MHD equations in the J-frame, converting quantities from one frame to the other when needed. We used an exponential coupling function, $\mathcal{A}\left( \chi \right) = \exp \left[ \alpha _0 \chi + \beta _0 \chi ^2 /2 \right]$, first introduced in \citet{damour_nonperturbative_1993}, in which the $\alpha _0$ parameter controls the weak field effects of the scalar field and $\beta _0$ regulates spontaneous scalarisation. The most stringent observational constraints, based on pulsar binaries, require that for massless scalar fields, $|\alpha _0| \lesssim 1.3 \times 10^{-3}$ and $\beta _0 \gtrsim -4.3$ (\citealt{voisin_2020}; see also \citealt{will_confrontation_2014} for a comprehensive review on tests of GR). For massive ones or for scalar fields endowed with a screening potential, lower values are still allowed \citep{doneva_rapidly_2016} as long as the screening radius is smaller than the binary separation. However, results found in a massless STT for the structure of NSs are also valid  for screened STTs as long as the screening radius is larger than the NS radius. This leaves open a large parameter space in terms of screening properties. We chose $\alpha _0=-2\times 10^{-4}$ and $\beta _0\in [-6,-4.5]$. By choosing this range of values, we want to highlight the effects of scalarisation while also showing its effects for values at the edge of the permitted parameter space for massless fields. For simplicity,  purely toroidal magnetic fields are computed assuming a magnetic barotropic law with a toroidal magnetisation index of $m=1$ (see \citetalias{soldateschi_2020} Eq.~47); whereas for purely poloidal magnetic fields, we opted for the simplest choice of a magnetisation function linearly dependent on the vector potential (see \citetalias{soldateschi_2020}  Eq.~43). We briefly recall here that the only known formalism to compute equilibria (even magnetised ones) in  the full non-linear regime, beyond the first order linear perturbation theory and beyond the Cowling approximation, is through the use of the generalised Bernoulli integral, including the case of differentially rotating stars, where the rotation rate is taken to be a function of the specific angular momentum \citep{bocquet_rotating_1995,kiuchi_relativistic_2008,frieben_equilibrium_2012,iosif_2014,pili_general_2017}, or through mathematically equivalent approaches. This sets severe constraints on the possible distribution of currents and, thereby, on the possible geometry of the magnetic field (the full functional dependence of the current density distribution can be found in \citetalias{soldateschi_2020}). For example, in the case of poloidal fields, the configuration is always dominated by the dipole term, but also contains  higher order multipoles. Our models have no surface currents. Typically, models with surface currents are not in true equilibria because they neglect the associated surface Lorentz force.
 
\begin{figure*}
    \centering
    \includegraphics[width=0.43\textwidth]{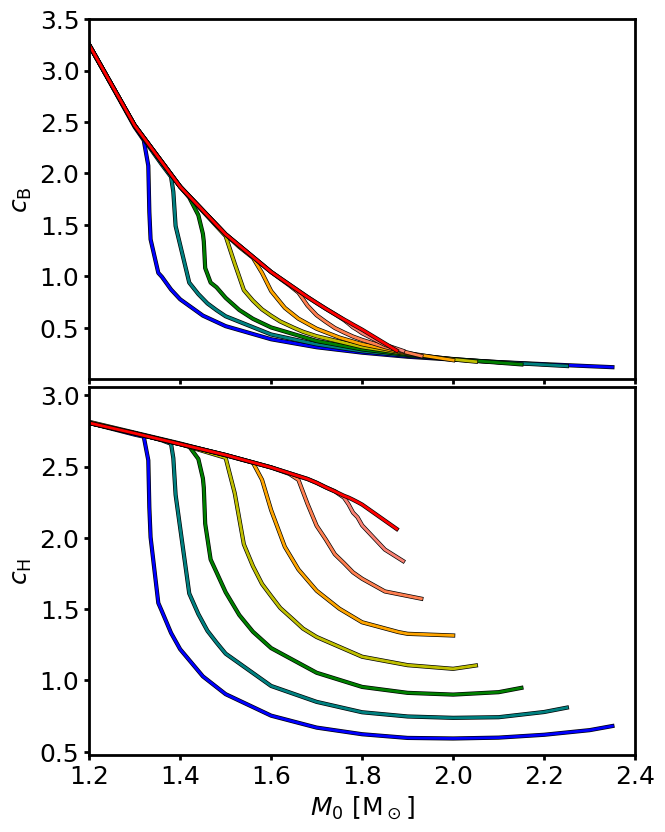}
        \includegraphics[width=0.43\textwidth]{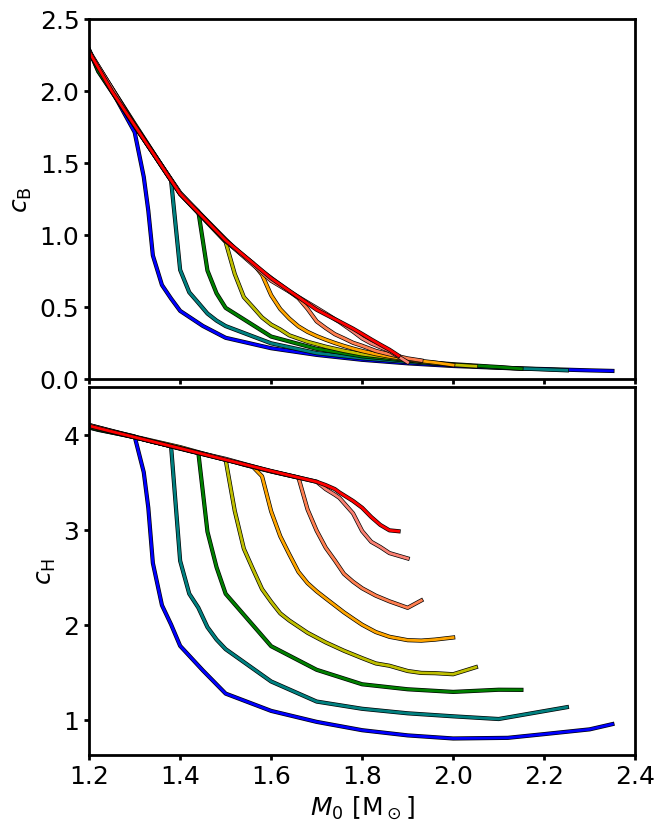}
    \caption{Distortion coefficients $c_\mathrm{B}$ (top panels) and $c_\mathrm{H}$ (bottom panels) as functions of the baryonic mass $M_0$ of models with a purely toroidal magnetic field (left panels) and with a purely poloidal magnetic field (right panels), for various value of $\beta _0$: from $\beta _0=-6$ (blue curve)  to $\beta _0=-4.5$ (light red curve) increasing by 0.25 with every line. The red curve corresponds to GR.}
    \label{fig:cbch}
\end{figure*}

We decided to parametrise the solution as a function of the baryonic mass $M_0$, which is the same in the E and J-frames. The relation with the E-frame Komar mass $\bar M$ is $\bar{M} \approx M_0-cM^2_0$, with $c=0.04$ $(0.05)$ for purely toroidal (poloidal) magnetic fields, and is the same in STT and GR. The behaviour of $c_\mathrm{B}$ and $c_\mathrm{H}$ as functions of $M_0$, for various $\beta _0$, are shown in Fig.~\ref{fig:cbch}, for NSs endowed with a purely toroidal or a purely poloidal magnetic field. The red line represents GR. The other lines represent the cases with a decreasing $\beta _0$, starting with $\beta _0=-4.5$ and going down to $\beta _0=-6$. We note that more scalarised sequences reach higher masses than less scalarised ones, because one of the effects of scalarisation is to increase the maximum possible mass of a stable NS, so that only heavily scalarised sequences are able to reach a baryonic mass as high as $\approx$~2.4~M$_\odot$ with our equation of state. The effect of scalarisation is clearly visible due to the distinctive rapid variation in the slope as the scalarised sequences depart from the GR one.  Decreasing the value of $\beta _0$ has the effect of enhancing the modifications with respect to GR and enlarging the scalarisation range. At a fixed $M_0$, scalarised NSs have a lower distortion coefficient - and a lower quadrupole deformation - than the corresponding GR models for most of the scalarisation range. In moving towards masses close to the maximum, the difference becomes increasingly small until it changes sign at the very end of the GR sequence.
\\\\
From a more quantitative point of view, the maximum relative difference between $c_\mathrm{B}$ in GR and in STT in the purely toroidal case is roughly 63\% for $\beta _0=-6$ and $M_0\approx 1.5$M$_\odot$. This difference decreases approaching 0 as $\beta _0$ increases. Moreover, as the baryonic mass increases, we can see that all sequences tend to coincide and reconnect to the GR one as the scalarisation range ends. As for $c_\mathrm{H}$, its maximum difference in STT relative to GR is roughly 72\% at $M_0\approx 1.8$M$_\odot$, for $\beta _0=-6$. Again, this difference decreases as $\beta _0$ increases. We note, however, that the various sequences of $c_\mathrm{H}$ do not seem to be reconnecting as the scalarisation range ends. This behaviour is to be attributed to the fact that the ratio $\tilde{\mathcal{H}}/\bar{W}$ depends on $M_0$, and, as such, it too exhibits the effect of scalarisation. In other words $\bar{W}$, at a fixed $M_0$, depends on $\beta _0$, which implies that in STTs $c_\mathrm{H}$ as defined in Eq.~\ref{eq:coeffs} is not directly comparable to GR at the same $\mathcal{H}$: first, it is needed to factor out the dependence of $\tilde{\mathcal{H}}/\bar{W}$ on $M_0$ and add it to $c_\mathrm{H}$. The same holds for purely poloidal magnetic fields. The maximum relative difference of $c_\mathrm{B}$ with respect to GR is 70\% for $\beta _0=-6$ at $M_0\approx 1.5$M$_\odot$, while for $c_\mathrm{H}$ it is 72\% at $M_0\approx 1.8$M$_\odot$ for $\beta _0=-6$.
\\\\
As we have seen, the distortion coefficients in STTs depart from the GR ones in a non-trivial way. Interestingly it looks like, apart from a scaling factor, both $c_\mathrm{H}$ and $c_\mathrm{B}$ have the same trend for toroidal and poloidal magnetic fields. In the case of $c_\mathrm{B}$, as can be seen from Fig.~\ref{fig:cbch}, for $M_0 < 1.6\mathrm{M}_\odot$, the values for toroidal magnetic fields are about a factor 1.5 higher than the cases with poloidal magnetic field. However, at higher masses, the trends are no longer similar between the two cases. Nonetheless, we have found that in the full range $1.2 \leq M_0/\mathrm{M}_\odot \lesssim 2.4 $ and $-6 \leq \beta _0 \leq -4.5 $, they are well approximated (to a few percents precision everywhere, except for the small range of masses in which scalarisation is triggered, where the error can reach a few tens of percents) by a combination of power laws of three global quantities defined for the corresponding unmagnetised model: the baryonic mass $M_0$, the J-frame circumferential radius $\tilde{R}_\mathrm{c}$ (see Eq.~C.8 in \citetalias{soldateschi_2020}) and the E-frame scalar charge $\bar{Q}_\mathrm{s}$ (see Eq.~C.7 in \citetalias{soldateschi_2020}). We note that these are not independent (for GR there is a one to one relation between mass and radius), but treating them as independent allows us to use simple power-law scalings in terms of global quantities.  In particular:
\begin{equation}\label{eq:approx}
        c_\mathrm{B}\approx c_1 M_{1.6}^\alpha R_{10}^\beta \left[ 1-c_2 Q_{1}^\gamma M_{1.6}^\delta R_{10}^\rho  \right] \; , 
\end{equation}
where the parameters are listed in Table~\ref{tab:coeffs}, $M_{1.6}$ is $M_0$ in units of $1.6 \mathrm{M}_\odot$, $R_{10}$ is $\tilde R _\mathrm{c}$ in units of $10\mathrm{km}$, and $Q_{1}$ is $\bar Q _\mathrm{s}$ in units of $1 \mathrm{M}_\odot$. The first term of Eq.~\ref{eq:approx} describes the distortion coefficient in GR, while the second term describes the deviation due to the presence of a scalar charge. First, for the GR term, we note that the coefficient $c_1$ of the models with toroidal field is about twice that of those with a poloidal one. The mass dependence is similar, while the exponent of the radius is higher by one for the poloidal field (this is likely due to the different geometry, prolate and oblate, of the configurations).  From the coefficients in Table~\ref{tab:coeffs}, we see that the second term of Eq.~\ref{eq:approx} has a more complex behaviour: the dependence on the scalar charge is similar, there is a weaker dependence on the mass for the poloidal field, while again the dependence on the radius is stronger by one power of $R_{10}$ in the poloidal case. The similarity between NSs with poloidal and toroidal magnetic fields is much stronger for $c_\mathrm{H}$, to the point that it is possible to derive a universal functional form over the entire mass range with an accuracy of few percents:
\begin{equation}\label{eq:chapprox}
        c_\mathrm{H}\approx 0.5+ \mathcal{F}(M_0)\mathcal{T}(M_0,\bar Q _\mathrm{s},\tilde R _\mathrm{c})\times\begin{cases} 0.65\;{\rm for\; toroidal}\\
        1.02\; {\rm for\; poloidal}\end{cases}\\
,\end{equation}
where $\mathcal{F}(M_0)$ represents the GR part and encodes the role of the equation of state, $\mathcal{T}(M_0,\bar Q _\mathrm{s},\tilde R _\mathrm{c})$ represents the correction due to scalarisation, and the oblate versus prolate geometry induced by the different magnetic field is encoded in the last factor. We find that: 
\begin{eqnarray}\label{eq:chterms}
&\mathcal{F}(M_0) &= 4.98-1.95 M_{1.6},\\
&\mathcal{T}(M_0,\bar{Q}_\mathrm{s},\tilde{R}_\mathrm{c})&=1-\frac{1.90}{R_{10}^{2.45}}\left(\frac{Q_1}{M_{1.6}}\right)^{1.3}
.\end{eqnarray}
It is well known that in GR, the coefficient $c_\mathrm{H}$ is only weakly dependent on the the mass, to the point that it can almost be taken as a constant. This is because the specific properties of the NS cancel out if the deformation is given as a function of $\mathcal{H}/W$. What we found here is that the same holds in STTs. The deformation is smaller than in GR, but the functional form of the correction is independent of the specific STT. Moreover, the geometry of the magnetic field is completely encoded in a constant coefficient that likely traces the oblate or prolate geometry of the star. 
\begin{table}
\caption{Values of the parameters for the approximations of $c_\mathrm{B}$ in Eq.~\ref{eq:approx} for purely toroidal and purely poloidal magnetic fields.}
\label{tab:coeffs}
\centering
\begin{tabular}{c c} 
 \hline\hline
 \noalign{\smallskip}
  Parameter & Toroidal \quad Poloidal  \\ [0.5ex]
  \noalign{\smallskip}
 \hline
  \noalign{\smallskip}
$c_1$ & 0.16 \quad 0.077 \\
$\alpha$ & -2.22 \quad -1.99 \\
$\beta$ & 4.86 \quad 5.80 \\
$c_2$ & 0.87 \quad 1.38 \\
$\gamma$ & 1.32 \quad 1.22 \\
$\delta$ & -1.27 \quad -0.86 \\
$\rho$ & -2.21 \quad -3.49 \\

 \noalign{\smallskip}
 \hline
\end{tabular}
\end{table}
\\\\
Since a quadrupolar deformation of the NS results in the emission of quadrupolar waves, both tensor and scalar, it is interesting to analyse what fraction of the energy contained in them is due to the quadrupole moment of the scalar field. For this purpose, we define the following ratios:
\begin{equation}\label{eq:ratios}
        \mathcal{S}= \bigg | \frac{q_\mathrm{s}}{q_\mathrm{g}} \bigg |\; ,       \; \mathcal{G}= \bigg | \frac{q_\mathrm{g}}{q^0_\mathrm{g}} \bigg | \; ,  
\end{equation}
where 
\begin{align}
        &q_\mathrm{s} = 2\pi \int \alpha _\mathrm{s} \mathcal{A}^4 \tilde{T}_\mathrm{p} \left( 3\sin ^2 \theta -2 \right) r^4 \sin \theta \mathrm{d}r \mathrm{d}\theta \; , \\
        &q_\mathrm{g} = \int \left[ \pi \mathcal{A}^4 (\tilde \varepsilon+\tilde \rho)-\frac{1}{8} (\partial \chi)^2 \right] r^4 \sin \theta \left(3\sin^2\theta -2 \right)  \mathrm{d}r \mathrm{d}\theta. 
\end{align}
These are, respectively, the Newtonian approximations of the `trace quadrupole' and of the `mass quadrupole' of the NS. The mass quadrupole $q_\mathrm{g}$ is just $\bar{I}_\mathrm{zz}-\bar{I}_\mathrm{xx}=\bar{e} \bar{I}_\mathrm{zz}$ (see Eq.~\ref{eq:deform}). We found that, in the mass range we investigated, $\bar{I}_\mathrm{zz}$ ranges from $6\times 10^{44}$g cm$^2$ to $4\times 10^{44}$g cm$^2$. This is consistent with GR, where the moment of inertia weakly depends on the mass \citep{lattimer_2001}, showing that the quadrupole is primarily encoded in the parameter $\bar{e}$. The quantities $\tilde \varepsilon$ and $\tilde \rho$ are respectively the J-frame internal energy density and rest-mass density, while  $(\partial \chi)^2= (\partial _r \chi)+r^{-2}(\partial _\theta \chi)^2$, and $q^0_\mathrm{g}$ is $q_\mathrm{g}$ calculated in GR. The quadrupole $q_\mathrm{s}$ is closely related to the `quadrupolar deformation of the trace' (see \citetalias{soldateschi_2020}, Eq.~C.19), which acts as the source of scalar waves. We note that these scalar waves are of a quadrupolar nature and differ from standard scalar monopolar GWs, which we do not consider here. In fact, a monopolar scalar wave, being monopoles rotationally invariant, can only arise following time-dependent monopolar variations of the structure of the NS
(e.g. when the star collapses, \citealt{gerosa_numerical_2016}) and is not triggered by the rotation of deformed NSs, to which our present results apply. This does not mean that rotation plays no role in monopolar waves emission since the vibrating eigenmodes depend on the NS structure, which also reflects  the underlying rotational profile. We note that the distinction between monopolar and quadrupolar waves depends only on the energy distribution of the waves (the multipolar pattern of the radiation), while the distinction between scalar and tensor modes depends on the nature of the waves (the spin of the wave carriers). The quantity of $\mathcal{S}$ gives a measure of which fraction of the energy lost in quadrupolar waves is contained in scalar modes, compared against tensor modes, while $\mathcal{G}$ quantifies the ratio of the energy of tensor modes in STTs versus GR. We note that the tensor GW luminosity scales approximately with $\bar{e}^2$, while it is the strain amplitude that scales with $\bar{e}$, so $\mathcal{S}$ and $\mathcal{G}$ are actually a measure of the variation in the strain, and it is their square to be related to the variation in the energy loss. It is worth pointing out that these ratios can be calculated by keeping fixed either $\tilde{B}_\mathrm{max}$ or $\tilde{\mathcal{H}}/\bar{W}$ and that unlike $\tilde{B}_\mathrm{max}$, $\tilde{\mathcal{H}}/\bar{W}$ depends on $M_0$ through $\bar W$ in different ways for different theories of gravity. Let's call this dependence $f(M_0,\beta _0)$ in our case, where the difference between STT and GR is encoded only by a varying $\beta _0$. This means that, in general, computing the ratios keeping fixed these two quantities does not yield the same result. In particular, ratios of quantities calculated with respect to models with the same $\beta _0$, like $\mathcal{S}$, are exactly the same in the two cases; instead, the ratio of a quantity in STTs over a quantity in GR, like $\mathcal{G}$, differ by a factor $f(M_0,\beta _0)/f(M_0,0)$.

\begin{figure*}
        \centering
        \includegraphics[width=0.43\textwidth]{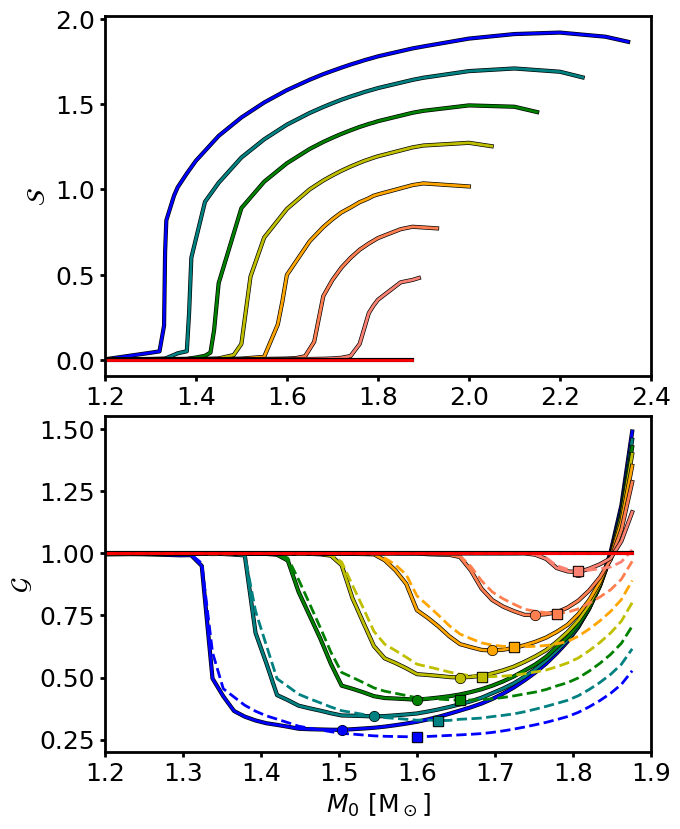}
        \includegraphics[width=0.43\textwidth]{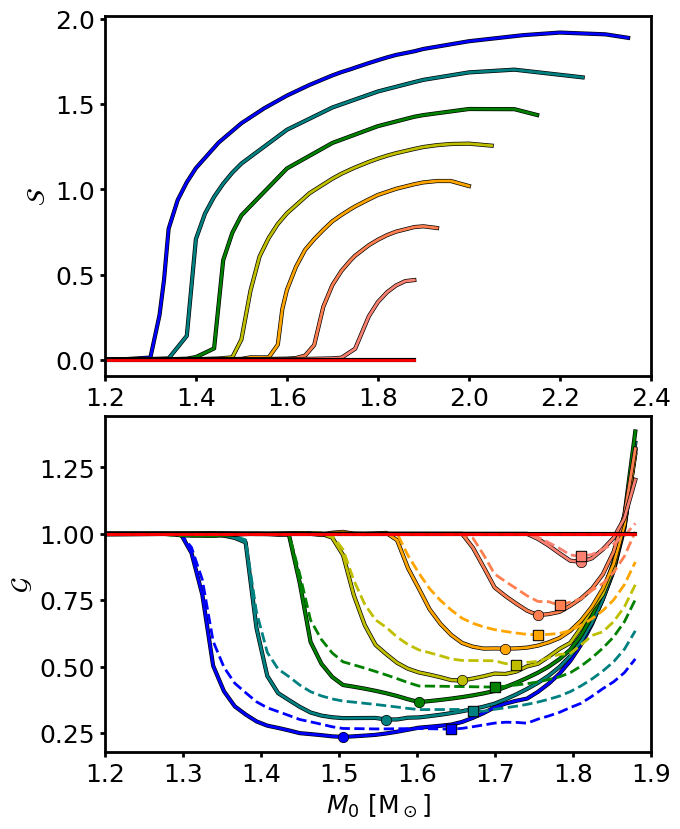}
    \caption{Ratios $\mathcal{S}$ (top panels) and $\mathcal{G}$ (bottom panels) as functions of the baryonic mass $M_0$ of models with a purely toroidal magnetic field (left panels) and with a purely poloidal magnetic field (right panels), for different value of $\beta _0$: from $\beta _0=-6$ (blue curve)  up to $\beta _0=-4.5$ (light red curve) increasing by 0.25 with every line. The red line corresponds to GR, where $\mathcal{S}=0$ and $\mathcal{G}=1$. Solid lines are the ratios computed by keeping $\tilde{B}_\mathrm{max}$ fixed, while dashed lines are obtained by keeping $\tilde{\mathcal{H}}/\bar{W}$ fixed. Markers show the models with minimum $\mathcal{G}$: circles for solid lines and squares for dashed lines.}
    \label{fig:ratios}
\end{figure*}

In Fig.~\ref{fig:ratios}, we show the ratios in Eq.~\ref{eq:ratios} for NSs endowed with a purely toroidal and a purely poloidal magnetic field, respectively. The top panels show that, when scalarisation occurs and $\mathcal{S}$ departs from zero, very rapidly $q_\mathrm{s} > q_\mathrm{g}$ for $\beta _0 \lesssim -5$, while sequences with $\beta _0 \gtrsim -5$ do not reach $\mathcal{S}=1$. This means that heavily scalarised NSs, once scalarisation kicks in, are dominated by losses due to scalar  radiation, while less scalarised ones are always dominated by tensor radiation. Differences in $\mathcal{S}$ between the purely toroidal and the purely poloidal cases are minimal, both in the scalarisation range and in the entity of its effect. The bottom panels of Fig.~\ref{fig:ratios} show that once scalarisation is triggered, $\mathcal{G}$ quickly falls down towards very low values; then, for NSs with a higher baryonic mass, $\mathcal{G}$ rises again very steeply. At a fixed $M_0$ below a threshold mass, the more scalarised the NS, the less energy is lost in tensor waves with respect to GR.  On the other hand, the more scalarised the NS, the more energy is injected in the scalar mode channel, as shown before with the $\mathcal{S}$ ratio, which increases almost monotonically with $M_0$. However, more massive NSs have a quadrupole deformation that is closer to GR than less massive stars, which reflects in the rise of $\mathcal{G}$ at high $M_0$. For masses close to the maximum, tensor waves losses can be even higher than in GR. It is interesting to note that all sequences intersect the line $\mathcal{G}=1$ at the same threshold mass $M_0 \approx 1.85$M$_\odot$. The reduction of the tensor GW strain is as high as 70\% (75\%) for $M_0=1.50$M$_\odot$ $(1.50$M$_\odot )$ for purely toroidal (poloidal) magnetic fields, for $\beta _0=-6$, while it is roughly 10\% for $M_0=1.80$M$_\odot$ and $\beta _0=-4.5$, for either purely toroidal or purely poloidal magnetic fields (see the blue and light red circles in Fig.~\ref{fig:ratios}). The same panels also show the ratio $\mathcal{G}$ computed by keeping fixed $\tilde{\mathcal{H}}/\bar{W}$. We can see that, in this case, the drop and especially the subsequent rise are less steep; in the case of a purely poloidal magnetic field, $\mathcal{G}$ is almost saturated to a constant value before slightly rising. 

 As was done before with the distortion coefficient, we found that $\mathcal{S}$ is well approximated (to a few percents precision everywhere but in the small range of masses where scalarisation occurs and the steepening is too strong to be well described by a simple power law) by
 \begin{equation}\label{eq:sratio}
        \mathcal{S} \approx 1.7 \left(\frac{R_{10}}{M_{1.6}^2}\right)^{0.2} Q_{1}^{1.2}\; 
 \end{equation}
 for both the toroidal and the poloidal cases. This shows again that it is possible to find a unifying functional dependence, even for scalar modes. Also, the value of  $\mathcal{G}$ looks very similar for the poloidal and toroidal cases, except at the largest masses above 1.6M$_\odot$ (we note that being a ratio with respect to GR it can only be computed up to maximum GR mass). On the other hand, due to its more complex behaviour, we did not find a satisfying approximation for $\mathcal{G}$ based on power laws of the quantities $M_0, \bar{Q}_\mathrm{s}, \tilde{R}_\mathrm{c}$.
 
 It is evident that the power emitted in tensor modes by scalarised NSs is smaller, for masses below $\approx 1.85$M$_\odot$, than for the model in GR of the same mass and same equation of state, even if the minimum does not correspond to the configuration with the strongest scalar charge. The trend changes for higher masses, where the losses in tensor modes in STTs are higher than for the corresponding GR models. Given that the scalar modes also contribute to the total energy losses, we see that as the mass rises, we first find a regime at the beginning of scalarisation,where the total GW emission is suppressed with respect to GR, which is then followed by a regime that is closer to the maximum mass where, due to the scalar channel, losses might even be enhanced by a factor of between 2 and 3.

The minimum value of $\mathcal{G}$ (marked by the circles and squares in Fig.~\ref{fig:ratios}) scales quadratically, with $\beta _0$ at fixed  $\tilde{B}_\mathrm{max}$ and $\tilde{\mathcal{H}}/\bar{W}$:
\begin{equation}
        \min \left( \mathcal{G} \right)\big|_{\tilde{B}_\mathrm{max}} \!\!\!\approx \begin{cases}7.79 + 2.36 \beta _0 + 0.185 \beta _0^2{\rm\; for\; toro.}\\
        7.98 + 2.44 \beta _0 + 0.190 \beta _0^2{\rm\; for\; polo.}
        \end{cases}\!\!\!,
\end{equation}
and
\begin{equation}
        \min \left( \mathcal{G} \right)\big|_{\tilde{\mathcal{H}}/\bar{W}} \!\approx \begin{cases}6.91 + 2.0 \beta _0 + 0.150 \beta _0^2{\rm\; for\; toro.}\\
        6.34 + 1.8 \beta _0 + 0.130 \beta _0^2{\rm\; for\; polo.}
        \end{cases}\!\!\!\!\!.
\end{equation}
Analogously, the mass at which the minimum of $\mathcal{G}$ occurs scales linearly with $\beta _0$:
\begin{equation}
        M_{0,\mathrm{min}} \big|_{\tilde{B}_\mathrm{max}} \!\!\!\approx 2.74 + 0.21 \beta _0 \; \mathrm{for\; toro.\;and\;polo.}\;,
\end{equation}
and 
\begin{equation}
        M_{0,\mathrm{min}} \big|_{\tilde{\mathcal{H}}/\bar{W}} \!\approx \begin{cases}
        2.45 + 0.143 \beta _0 {\rm\; for\; toro.}\\
        2.32 + 0.113 \beta _0 {\rm\; for\; polo.}
        \end{cases}\!\!\!.
\end{equation}

\section{Conclusions}\label{sec:conclusions}
In this work we explore how the addition of a scalar field that is non-minimally coupled to the metric affects the magnetic quadrupolar deformation of a NS in the weak field regime ($\tilde{B}_\mathrm{max}\lesssim 10^{17}$G). We find, as in Newtonian gravity and in GR, in this limit the quadrupolar deformation, $\bar{e,}$ can be well-approximated by a linear function of either $\tilde{B}^2_\mathrm{max}$ or $\tilde{\mathcal{H}}/\bar{W}$, for fixed baryonic mass, $M_0,$ and scalarisation parameter, $\beta _0$. We find that the coefficients of the linear expansion strongly depart from those of GR for sufficiently negative values of $\beta _0$: for the range of parameters investigated here, spontaneous scalarisation can decrease the magnetic deformation of a NS by up to $\approx 70\%$ of the GR value, for $\beta _0 =-6$. For values of $\beta _0 \gtrsim -4.3,$ we find that the results in massless STTs without screening are indistinguishable from GR.
This behaviour can be attributed to the interplay between various effects. First, given a certain EoS, we find that NSs in STTs have a different central density than those in GR with the same mass. In particular, below a threshold mass in the stable branch of the mass-central density diagram, scalarised models have a higher central density than the GR models with the same mass, which causes the inner region of the star to be less prone to deformation; the opposite happens for masses above the threshold, which are more susceptible to deformation than the corresponding GR models with the same mass. This tendency explains why $\mathcal{G}$ is higher than unity for $M_0 \gtrsim 1.85$M$_\odot$, which is very close to the threshold mass, which we found to be $M_0 \approx 1.88$M$_\odot$ for low magnetisations. Moreover, NSs endowed with a purely toroidal magnetic field have a prolate shape, with a more pronounced density gradient at the equator than at the pole, which, in turn, generates a steeper gradient of the scalar field at the equator than at the pole. Given that the effective pressure of the scalar field depends on its spatial derivatives and has the same sign as the fluid pressure, we see that its effect is to reduce the deformation of the star, making it more spherical. The same qualitative behaviour is exhibited by a star endowed with a purely poloidal magnetic field and which possesses an oblate shape with a steeper density gradient at the pole, thus causing the scalar field to apply more pressure in the polar direction than in the equatorial one. As expected, a more pronounced scalarisation (i.e. a more negative $\beta _0$) reflects in a stronger pressure of the scalar field, rendering the star even more spherical, thus reducing the distortion coefficients further. It is known that the scalar field can act as a `stabiliser' for NSs, rendering their shape more spherical. Finally, the scalar field acts as an effective coupling term (it replaces the inverse of the gravitational constant of GR) between matter and the metric. In more scalarised systems, or in the NS central region where the scalar field is larger, this coupling is weaker, and this also holds for perturbations of the energy momentum tensor. Thus, the same structural deformation of the NS produces a weaker metric deformation. All these effects depend on where the deformation is located (centre versus the outer layers).

It is interesting to note that unlike the quadrupolar deformation of NSs caused by their rotation (see e.g. \citealt{doneva_iq_2014}), the magnetic quadrupolar deformation, as we have seen, decreases in STTs with respect to GR, except for masses close to the maximum. This difference can be explained by the fact that rotation, unlike a magnetic field, affects mostly the outer layers of the NS, which in scalarised systems are less gravitationally bound than in GR (scalarized NS have larger radii than in GR), increasing their deformability with respect to GR. Magnetic deformation seems to be instead mostly regulated by the density in the central region where the magnetic field strength peaks. 

Regarding GWs, STTs predict the existence of scalar waves, which, as we show here, can have an amplitude comparable to standard tensor waves and might even dominate the GW losses for strongly scalarised NSs. This can lead to a point when the total emission is even larger than in GR.

We find that a good approximation of the distortion coefficients is given by a simple power-law dependence on $M_0$, $\tilde{R}_\mathrm{c}$ , and $\bar{Q}_\mathrm{s}$.
More interestingly, we found that in terms of the ratio of magnetic to binding energy the effect of a scalar field on the NS deformation and the ratio of scalar to tensor waves emission can be parametrised by a unique function independently of the magnetic field structure or of the STT parameters. It seems that the presence of a scalar charge can easily be factorised. This is a generalisation of what was already known for GR, that is, with $c_{\mathrm{H}}$ being almost a constant. As we described above, the magnetic deformability of NSs heavily depends on their internal structure, determined by the EoS. We leave to a future work to verify how much the functional form and the coefficients entering such function depend on the EoS. 
\\\\
Since the quadrupolar deformation of NSs, at a fixed $M_0$ and for most of the scalarisation range, is reduced in STTs with respect to GR, it is expected that the energy lost in tensor GWs by deformed NSs is also reduced in STTs; on the other hand, it is enhanced for masses close to the maximum mass for a stable NS in GR. In fact, we found that strongly scalarised stars with a baryonic mass around $1.5$M$_\odot$ have a tensor GW strain, $h_0,$ that is up to $75\%$ lower than for the corresponding stars in GR, while it is roughly $10\%$ lower, for $\beta _0=-4.5$, for masses of $1.8$M$_\odot$. This means that, for values of $\beta _0$ currently allowed by observations, a less than $10\%$ variation in $h_0$ is to be expected for massless scalar fields. This is much smaller than the typical uncertainties over the distances and the strength of magnetic fields, even for well-constrained galactic objects. Higher values might hold for massive scalar fields and this could lead to serious underestimations (or overestimations, depending on the mass) of the energetics of the system and all that follows from that, such as its distance or the strength of its magnetic field. If similar results on the role of the scalar field in the modification of tensor modes hold as well for other kinds of deformation (e.g. tidal deformations of scalarised NSs in mergers), this could have a deep impact on our understanding of binary NS merger events \citep{abbott_gw170817:_2017}.
More interestingly, we found that the scalar mode can be emitted carrying an energy comparable to the tensor one. However, its strain is suppressed by a factor $\alpha _0 \sim 10^{-5} - 10^{-4}$, which weakens the coupling of the scalar mode to the detector and renders the possibility of it being detected even fainter.

Our results are computed in the full non-linear regime. We also computed the quadrupole deformation, holding  both the metric and the scalar field fixed, which can be thought of as a Cowling approximation in STTs. In this case we found that for the mass range we investigated, the coefficients $c_\mathrm{B}$ and $c_\mathrm{H}$ are smaller by a factor $\approx 0.5-0.65$.

The current sensitivities of the LIGO-Virgo observatories \citep{ligo_virgo_kagra_2019} could be enough to detect tensor CGWs emitted by galactic neutrons stars remnants form merger events, with  millisecond period (lasting few seconds), if the quadrupole deformation is $e \gtrsim  10^{-5}$ \citep{lasky_2015,ligo_ellipticity_2020}. Future GW detector of the class of Einstein Telescope \citep{punturo_2010} and Cosmic Explorer \citep{reitze_2019} could detect deformations as low as $e \gtrsim  10^{-6}$. This means that, for what concerns continuous scalar waves, detectors with same sensitivity could reveal them from millisecond NSs, spinning for few seconds, only if the scalar quadrupole is $\alpha _0^{-1}$ time bigger, which means magnetic field strength of the order of few $10^{17}$G, at the limit of the values that can be reached \citep{ciolfi_2019}. For the same magnetic fields, tensor waves could be much more easily detected. Instead, the detection of scalar CGWs from slowly spinning magnetars with internal fields of few $10^{17}$G \citep{olausen_2014,frederick_2020} requires instruments with a sensitivity at least one order of magnitude better than DECIGO \citep{kawamura_2008} and BBO \citep{harry_2006}.

We caution the reader that there is evidence that the magnetic field at the surface or in the magnetosphere of NSs can have strong multipoles \citep{bignami_2003,nicer_2019,braking_index_2020,raynaud_2020}. However, the interpretation of the data is not unambiguous (e.g. the magnetic field inferred from cyclotron lines changes if   either electron or proton cyclotron are assumed). It is even less clear how these apply to the magnetic field in the interior, whose geometry is totally unknown. From a theoretical point of view, it is reasonable to expect a difference between the interior and surface magnetic fields. The evolution of the latter is mostly dictated by the Hall term associated to crustal impurities, which leads to the formation of small-scale structures and higher order multipoles \citep{pons_2019,de_grandis_2020}, while the dissipation of the former is mostly Ohmic, preferentially suppressing small-scale structure and higher multipoles \citep{haensel_1990}. Given that here we are mostly interested in
the deviation with respect to GR, we can consider the purely toroidal
and purely poloidal cases as two extrema of the much larger space of possible
magnetic configurations. Our results showing that the deviations with respect to
GR due to a scalar field are very similar in these two extrema lend us confidence to the consideration
that similar deviations with respect to GR will also apply in more complex
magnetic field geometries \citep{mastrano_2013,mastrano_2015}.


\begin{acknowledgements}
The authors also acknowledge financial support from the ``Accordo Attuativo ASI-INAF n. 2017-14-H.0 Progetto: on the escape of cosmic rays and their impact on the background plasma'' and from the INFN Teongrav collaboration.
\end{acknowledgements}



\bibliographystyle{aa}
\bibliography{articolo.bib} 


%
%
%


\label{lastpage}
\end{document}